\documentclass[11pt,eqs]{article}
\usepackage{latexsym}
\usepackage{graphicx}
\usepackage{bbm}

\makeatletter
\newcommand\xleftrightarrow[2][]{%
  \ext@arrow 9999{\longleftrightarrowfill@}{#1}{#2}}
\newcommand\longleftrightarrowfill@{%
  \arrowfill@\leftarrow\relbar\rightarrow}
\makeatother

\title{
Simultaneous T-dualization of type II pure spinor superstring
\thanks{Work supported in part by
the Serbian Ministry of Education, Science and Technological Development, under contract No. 171031.}}
\author{B. Nikoli\'c and B. Sazdovi\'c
\thanks{e-mail: bnikolic, sazdovic@ipb.ac.rs}\\
{\it Institute of Physics Belgrade,}\\
{\it University of Belgrade,}\\
{\it Pregrevica 118, Serbia}}

\begin{document}

\maketitle
\begin{abstract}
In this article we consider simultaneous T-dualization of type II superstring action in pure spinor formulation. Simultaneous T-dualization
means that we make T-dualization at the same time along some subset of initial coordinates marked by $x^a$. The only imposed assumption stems from 
the applicability of the Buscher T-dualization procedure -
background fields do not depend on dualized directions $x^a$. In this way we obtain the full form of the T-dual background fields and T-dual transformation laws. 
Because two chiral sectors transform differently, there are two sets of vielbeins and gamma matrices connected by the local Lorentz transformation. Its spinorial representation is the same as in the constant
background case. We also found the full expression for T-dual dilaton field.
\end{abstract}

\section{Introduction}
\setcounter{equation}{0}

The importance of T-duality rose after M-theory was discovered. Five consistent superstring theories are connected by web of T and S dualities and make M-theory \cite{mtheory}. 
For example, T-duality connects type IIA and type IIB superstring
theories in the sense that after odd number of T-dualizations type IIA/B turns into IIB/A, while after even number of T-dualizations type IIA/B stays unchanged \cite{IIAIIB,BNBSCBC}.

T-dualization of type II superstrings was a subject of the papers \cite{GR,H,BPT,BNBSCBC}. In some articles T-dualization along single direction is considered \cite{H,BPT}. 
Two chirality sectors transform under T-duality differently and, consequently, in T-dual picture
there are two sets of vielbeins and gamma matrices. But there is a local Lorentz transformation connecting them. 
In Refs.\cite{H,BPT}, in the case of T-dualization along one direction, spinorial representation of that local Lorentz transformation is found. Detailed derivation is presented in Ref.\cite{H}.

The case of simultaneous T-dualization of pure spinor type II superstring theory is investigated in Ref.\cite{BNBSCBC}.
By simultaneous T-dualization we denote T-dualization along some subset of coordinates at the same time \cite{BNBSCBC,erik}.
In Ref.\cite{BNBSCBC} we used the action in the approximation of
constant background fields obtained from general one given in Ref.\cite{verteks} after some assumptions. First, we took that all background fields are $x^\mu$ independent
justifying such assumption with the possibility of making simultaneous T-dualization along any subset of coordinates including full T-dualization. 
The second crucial assumptions was technical one. Because the full action of Ref.\cite{verteks}
is in the form of an expansion in powers of $\theta^\alpha$ and $\bar\theta^\alpha$, for technical simplicity, we took into consideration only basic terms which are $\theta^\alpha$ and $\bar\theta^\alpha$
independent. Effectively, only physical superfields (their first components are identified with supergravity fields) survive and they are constant. 
Using obtained action, in Ref.\cite{BNBSCBC} we investigated simultaneous T-dualization 
and obtained the transformation laws connecting initial
and T-dual coordinates and the expressions for T-dual background fields. We presented a detailed derivation of local Lorentz transformation in the spinorial representation.
Also we discussed the case of timelike T-dualization and prove the results of Ref.\cite{timelike} obtained in the analysis of effective action.

Mathematical framework for T-dualization is developed by Buscher \cite{B}. Standard Buscher T-dualization procedure is applicable if the theory has shift symmetry. 
This means that it is possible to find such coordinate basis in which background fields do not depend on some directions \cite{B,S,RV,GPR,AABL}. Localization of the symmetry is done in a standard way - by replacing 
the world-sheet
derivatives $\partial_\pm x^\mu$ with covariant ones, $D_\pm x^\mu=\partial_\pm x^\mu+v_\pm^\mu$, where $v_\pm^\mu$ are gauge fields. 
In order to
make T-dual theory physically equivalent to the initial one, a term with Lagrange multiplier is added to the action so that the field strength should be zero. As a consequence of the shift 
symmetry, we can fix initial coordinates and obtain so called gauge fixed action. Variation of this action with respect to
the Lagrange multiplier produces initial action, while variation with respect to the gauge fields gives T-dual action. When one applies the procedure in the open string case \cite{open} then one has to consider both equations of motion and 
boundary conditions. Consequently, $Dp$-branes appear in the analysis.

There are two main approaches in superstring theory - Neveu-Schwarz-Ramond (NSR) and Green-Schwarz (GS) formalism \cite{S}. The first one contains explicit world-sheet supersymmetry, while the second one has explicit
space-time supersymmetry. There are some disadvantages of these approaches: in NSR formalism Ramond-Ramond (R-R) sector is missing and space-time supersymmetry
is not manifest, while in GS formalism quantization can be performed just in light-cone gauge. In last two decades a new approach has appeared, pure spinor formalism \cite{berko,pspin,susyNC,NPBref,verteks}.
It is pretty similar to GS one in the sense that space-time supersymmetry is manifest but it contains pure spinors $\lambda^\alpha$ and $\bar\lambda^\alpha$ satisfying so called pure spinor constraints,
$\lambda^\alpha (\Gamma^\mu)_{\alpha\beta}\lambda^\beta=\bar\lambda^\alpha(\Gamma^\mu)_{\alpha\beta}\bar\lambda^\beta=0$. Pure spinor formalism uses advantages of the previous two
formalisms and avoids some disadvantages. In this article we will use pure spinor action of type II superstring from Ref.\cite{verteks}, where detailed derivation of the action is presented. The action is given in the form
of an expansion in powers of $\theta^\alpha$ and $\bar\theta^\alpha$ obtained using (anti)holomorphicity and nilpotency conditions.

In this article we study simultaneous T-dualization of the pure spinor superstring type II theory with only one assumption - background fields are independent of
the coordinates $x^a$ along which we make T-dualization. This assumption stems from the applicability of the Buscher procedure. 
Our main goal is to find the full form of all
T-dual background fields and 
T-dual transformation laws.

We start with the action (\ref{eq:vsg}) and decompose the variables $X^M$ and $\bar X^M$ (\ref{eq:strujejpmM}) extracting directions $x^a$ along which we make T-dualization. 
Then we perform
Buscher T-dualization procedure along $x^a$ obtaining the T-dual transformation laws and T-dual action. Two chirality sectors transform differently under T-dualization. 
Consequently, there are two sets of vielbeins and gamma matrices,
which are connected by local Lorentz transformation represented by matrix ${}_a \Omega$.
In order to work with unique set of gamma matrices, we introduce proper fermionic variables keeping unbar fermionic variables unchanged while bar variables are corrected by matrix ${}_a \Omega$. After introducing proper fermionic variables, we read the full form of the T-dual background fields.

We get the explicit expressions for T-dual physical superfields. For constant background they turn to the result of Ref.\cite{BNBSCBC}. Expressions for the auxiliary superfields and 
field strengths are completely new in the sense that they are missed in the constant background case. In order to avoid long expressions, we will give explicitly just expression for T-dual auxiliary field ${}_a A^a{}_\alpha$ (\ref{eq:uvod1}) and expression for the field strength ${}_a \Omega^{a,\hat\mu\hat\nu}$ (\ref{eq:uvod2}).

Dilaton field is treated within quantum formalism. We obtain the most general expression for T-dual dilaton field within pure spinor formulation of type II superstring theory.


\section{Type II pure spinor superstring theory}
\setcounter{equation}{0}

In this section we will introduce the type II pure spinor superstring action in compact and expanded form.

The sigma model action for type II superstring of
Ref.\cite{verteks} is of the form
\begin{equation}\label{eq:vsg}
S=\int_\Sigma d^2 \xi (X^T)^M A_{MN}\bar X^N+S_\lambda+S_{\bar\lambda}\, ,
\end{equation}
where $X^M$ and $\bar X^N$ are left and right chiral supersymmetric variables
\begin{equation}\label{eq:XMXM}
X^M=\left(
\begin{array}{c}
\partial_+ \theta^\alpha\\\Pi^\mu\\d_\alpha\\\frac{1}{2}N^{\mu\nu}
\end{array}\right)\, ,\quad \bar X^M=\left(
\begin{array}{c}
\partial_-\bar\theta^\alpha\\\bar\Pi^\mu\\\bar d_\alpha\\\frac{1}{2}\bar N^{\mu\nu}
\end{array}\right),\quad \left[M=(\alpha, \mu, \alpha, \mu\nu)\right]\, ,
\end{equation}
which components are defined as
\begin{equation}
\Pi^\mu=\partial_+ x^\mu+\frac{1}{2}\theta^\alpha (\Gamma^\mu)_{\alpha\beta}\partial_+ \theta^\beta\, ,\quad \bar\Pi^\mu=\partial_- x^\mu+\frac{1}{2}\bar\theta^\alpha (\Gamma^\mu)_{\alpha\beta}\partial_- \bar\theta^\beta\, ,
\end{equation}
\begin{eqnarray}
d_\alpha&=&\pi_\alpha-\frac{1}{2}(\Gamma_\mu \theta)_\alpha\left[ \partial_+ x^\mu +\frac{1}{4} (\theta \Gamma_\mu \partial_+\theta)\right]\, ,\nonumber\\
\bar d_\alpha&=&\bar\pi_\alpha-\frac{1}{2}(\Gamma_\mu \bar\theta)_\alpha \left[\partial_- x^\mu +\frac{1}{4} (\bar\theta \Gamma_\mu \partial_-\bar\theta)\right]\, ,
\end{eqnarray}
\begin{equation}\label{eq:Npm}
N^{\mu\nu}=\frac{1}{2}w_\alpha(\Gamma^{[\mu\nu]})^\alpha{}_\beta \lambda^\beta\, ,\quad \bar N^{\mu\nu}=\frac{1}{2}\bar w_\alpha (\Gamma^{[\mu\nu]})^\alpha{}_\beta \bar\lambda^\beta\, .
\end{equation}
In the analysis we will use the action in the form (\ref{eq:vsg}). Just for completeness, the expanded form of the action is
\begin{eqnarray}\label{eq:VSG}
S&=&\int d^2\xi \left[\partial_+ \theta^\alpha A_{\alpha\beta}\partial_-\bar\theta^\beta+\partial_+ \theta^\alpha A_{\alpha\mu}\bar\Pi^\mu+\Pi^\mu A_{\mu\alpha}\partial_-\bar\theta^\alpha+\Pi^\mu A_{\mu\nu}\bar\Pi^\nu\right.\nonumber\\
&+& d_\alpha E^\alpha{}_\beta \partial_-\bar\theta^\beta+d_\alpha E^\alpha{}_\mu \bar\Pi^\mu+\partial_+ \theta^\alpha E_\alpha{}^\beta \bar d_\beta+\Pi^\mu \bar E_\mu{}^\beta \bar d_\beta+d_\alpha {\rm P}^{\alpha\beta}\bar d_\beta\nonumber\\
&+& \frac{1}{2}N^{\mu\nu} \Omega_{\mu\nu,\beta} \partial_- \bar\theta^\beta+\frac{1}{2}N^{\mu\nu} \Omega_{\mu\nu,\rho} \bar\Pi^\rho+\frac{1}{2}\partial_+\theta^\alpha \Omega_{\alpha,\mu\nu}\bar N^{\mu\nu}+\frac{1}{2}\Pi^\mu \Omega_{\mu,\nu\rho}\bar N^{\nu\rho}\nonumber\\
&+&\left. \frac{1}{2}N^{\mu\nu}\bar C_{\mu\nu}{}^\beta \bar d_\beta+\frac{1}{2}d_\alpha C^\alpha{}_{\mu\nu} \bar N^{\mu\nu}+\frac{1}{4}N^{\mu\nu} S_{\mu\nu,\rho\sigma}\bar N^{\rho\sigma}\right]+S_\lambda+S_{\bar\lambda}\, ,
\end{eqnarray}
where we used matrix $A_{MN}$ in explicit form
\begin{equation}\label{eq:Amn}
A_{MN}=\left(\begin{array}{cccc}
A_{\alpha\beta} & A_{\alpha\nu} & E_\alpha{}^\beta & \Omega_{\alpha,\mu\nu}\\
A_{\mu\beta} & A_{\mu\nu} & \bar E_\mu{}^\beta & \Omega_{\mu,\nu\rho}\\
E^\alpha{}_\beta & E^\alpha{}_\nu & {\rm P}^{\alpha\beta} & C^\alpha{}_{\mu\nu}\\
\Omega_{\mu\nu,\beta} & \Omega_{\mu\nu,\rho} & \bar C_{\mu\nu}{}^\beta & S_{\mu\nu,\rho\sigma}
\end{array}\right)\, .
\end{equation}
Matrix $A_{MN}$, containing type II superfields, generally depends on $x^\mu$,
$\theta^\alpha$ and $\bar\theta^\alpha$.
The superfields $A_{\mu\nu}$, $\bar E_\mu{}^\alpha$, $E^\alpha{}_\mu$ and ${\rm P}^{\alpha\beta}$ are physical superfields, because their first components are supergravity fields.
The fields in the first column and first row are auxiliary superfields because they can be expressed in
terms of the physical ones \cite{verteks}. The rest ones, $\Omega_{\alpha,\mu\nu}$, $\Omega_{\mu\nu,\alpha}$, $\Omega_{\mu,\nu\rho}(\Omega_{\mu\nu,\rho})$, $C^\alpha{}_{\mu\nu}(\bar C_{\mu\nu}{}^\alpha)$ and $S_{\mu\nu,\rho\sigma}$, are curvatures (field strengths).

The world sheet $\Sigma$ is parameterized by
$\xi^m=(\xi^0=\tau\, ,\xi^1=\sigma)$ and
$\partial_\pm=\partial_\tau\pm\partial_\sigma$. Superspace is spanned by bosonic coordinates $x^\mu$ ($\mu=0,1,2,\dots,9$) and fermionic ones $\theta^\alpha$ and $\bar\theta^{\alpha}$
$(\alpha=1,2,\dots,16)$. The variables $\pi_\alpha$ and $\bar
\pi_{\alpha}$ are canonically conjugated momenta to
$\theta^\alpha$ and $\bar\theta^\alpha$, respectively. The actions for pure spinors, $S_\lambda$ and $S_{\bar\lambda}$, are free field actions
\begin{equation}
S_\lambda=\int d^2\xi w_\alpha \partial_-\lambda^\alpha\, ,\quad S_{\bar\lambda}=\int d^2\xi \bar w_\alpha \partial_+ \bar\lambda^\alpha\, ,
\end{equation}
where $\lambda^\alpha$ and $\bar\lambda^\alpha$ are pure spinors and $w_\alpha$ and $\bar w_\alpha$ are their canonically conjugated momenta, respectively. The pure spinors satisfy so called pure spinor constraints
\begin{equation}\label{eq:psc0}
\lambda^\alpha (\Gamma^\mu)_{\alpha\beta}\lambda^\beta=\bar\lambda^\alpha (\Gamma^\mu)_{\alpha\beta}\bar\lambda^\beta=0\, .
\end{equation}

We are going to perform T-dualization along some subset of bosonic coordinates $x^a$. So, we will assume that these directions are Killing vectors and that background fields do not depend on them.


\section{T-dualization along arbitrary number of coordinates}
\setcounter{equation}{0}

In this section we will make T-dualization along arbitrary subset of the coordinates $x^a$. First we will make mathematical preparation extracting the desired directions from variables
$X^M$ and $\bar X^M$. Then we will apply standard Buscher procedure assuming that background fields do not depend on $x^a$. 

\subsection{Mathematical preparation}

In order to make T-dualization along arbitrary bosonic directions $x^a$, let us split the space time index $\mu$ in $a$ and the undualized ones, $i$. 
We write the variables $X^M$ and $\bar X^N$ in the appropriate form, separating derivatives of T-dualized coordinates $x^a$ 
\begin{eqnarray}\label{eq:strujejpmM}
X^M&=&P^M{}_{a} \partial_+ x^a+{}_aj^M_+\equiv P^M{}_{a} \partial_+ x^a+P^M{}_{i} \partial_+ x^i+j_+^M\, ,\nonumber \\ \bar X^M&=&\bar P^M{}_{a} \partial_- x^a+{}_a j_-^M \equiv\bar P^M{}_{a} \partial_- x^a+\bar P^M{}_{i} \partial_- x^i+j_-^M\, ,
\end{eqnarray}
where
\begin{equation}
P^M{}_{b}=\left(
\begin{array}{c}
0\\ \delta^a{}_b\\ 0 \\ -\frac{1}{2}(\Gamma_b \theta)_{\alpha}\\ 0
\end{array}\right)\, ,\quad \bar P^M{}_{b}=\left(
\begin{array}{c}
0\\ \delta^a{}_b\\ 0 \\ -\frac{1}{2}(\Gamma_b \bar\theta)_{\alpha}\\ 0
\end{array}\right)\, ,
\end{equation}
\begin{equation}
P^M{}_{j}=\left(
\begin{array}{c}
0\\ 0 \\\delta^i{}_j\\ -\frac{1}{2}(\Gamma_j \theta)_{\alpha}\\ 0
\end{array}\right)\, ,\quad \bar P^M{}_{j}=\left(
\begin{array}{c}
0\\ 0\\ \delta^i{}_j\\ -\frac{1}{2}(\Gamma_j \bar\theta)_{\alpha}\\ 0
\end{array}\right)\, ,
\end{equation}
\begin{equation}\label{eq:astrujap}
{}_a j^M_+=\left(
\begin{array}{c}
\partial_+ \theta^\alpha\\ \frac{1}{2}(\Gamma^a\theta)_\alpha \partial_+ \theta^\alpha\\ \Pi^i \\ \pi_\alpha-\frac{1}{2}(\Gamma_i\theta)_\alpha \partial_+ x^i-\frac{1}{8}(\Gamma^\mu\theta)_\alpha (\theta \Gamma_\mu \partial_+ \theta)\\ \frac{1}{2}N^{\mu\nu}
\end{array}\right)\, ,
\end{equation}
\begin{equation}\label{eq:astrujam}
{}_a j^M_-=\left(
\begin{array}{c}
\partial_- \bar\theta^\alpha\\ \frac{1}{2}(\Gamma^a\bar\theta)_\alpha \partial_- \bar\theta^\alpha\\ \bar\Pi^i \\ \bar\pi_\alpha-\frac{1}{2}(\Gamma_i\bar\theta)_\alpha \partial_- x^i-\frac{1}{8}(\Gamma^\mu\bar\theta)_\alpha (\bar\theta \Gamma_\mu \partial_- \bar\theta)\\ \frac{1}{2}\bar N^{\mu\nu}
\end{array}\right)\, .
\end{equation}
In comparison with (\ref{eq:XMXM}) we split $\Pi^\mu$ into $\Pi^a$ and $\Pi^i$ as well as
$\Gamma^\mu$ into $\Gamma^a$ and $\Gamma^i$. Consequently, variables $X^M$ and $\bar X^M$ have five block components and $A_{MN}$ is $5\times 5$ block matrix, where index $M=(\alpha,a,i,\alpha,\mu\nu)$.

Let us introduce the notation
\begin{equation}\label{eq:tildepiab2}
\tilde {\cal A}_{ab}=P_{a}{}^M A_{MN} \bar P^N{}_{b}=A_{ab}-\frac{1}{2}\bar E_a{}^\alpha (\Gamma_b \bar\theta)_\alpha-\frac{1}{2}(\Gamma_a \theta)_\alpha E^\alpha{}_b+\frac{1}{4}(\Gamma_a \theta)_\alpha P^{\alpha\beta}(\Gamma_b \bar\theta)_\beta\, ,
\end{equation}
\begin{equation}\label{eq:tildepiab}
\tilde \Pi_{+ab}\equiv \frac{1}{\kappa}\tilde{\cal A}_{ab}\, ,\quad\tilde \Pi_{\pm ab}=\tilde B_{ab}\pm\frac{1}{2}\tilde G_{ab}\, ,
\end{equation}
\begin{equation}\label{eq:strujeJ}
J_{+a}=\frac{2}{\kappa}{}_a j_+^M A_{MN} \bar P^N{}_a\, ,\quad J_{-a}=-\frac{2}{\kappa} P_a{}^M A_{MN} {}_a j^N_-\, .
\end{equation}
The field $\tilde B_{ab}$ plays a role of Kalb-Ramond field, and $\tilde G_{ab}$ the role of metric in the process of T-dualization.

Applying this decomposition to the action (\ref{eq:vsg}), it gets a form
\begin{equation}\label{eq:parcvsg}
S=\int d^2\xi \left( \kappa\partial_+ x^a \tilde \Pi_{+ab}\partial_- x^b-\frac{\kappa}{2}\partial_+ x^a J_{-a}+\frac{\kappa}{2}J_{+a}\partial_- x^a+{}_a j_{+}^M A_{MN}{}_a j^N_-\right)\, ,
\end{equation}
where all terms with derivatives $\partial_\pm x^a$ are written explicitly.

\subsection{Buscher procedure}

Let us perform T-dualization of the action (\ref{eq:parcvsg}) along $x^a$ directions. We assume that $x^a$ directions are Killing ones, so, background fields do not depend on them.
Applying standard procedure of Buscher T-dualization we replace ordinary world-sheet derivatives $\partial_\pm x^a$ by covariant ones
\begin{equation}
D_\pm x^a=\partial_\pm x^a+v^a_\pm\, .
\end{equation}
In order to make the fields $v^a_\pm$ unphysical we add the term
\begin{equation}
S_L=\frac{\kappa}{2}\int d^2\xi  ( v_+^a\partial_- y_a-v^a_-\partial_+ y_a)\, , 
\end{equation}
where $y_a$ are Lagrange multipliers. Taking into account that $x^a$ are isometry directions we choose the gauge, $x^a=0$, so that the gauge fixed action takes a form
\begin{eqnarray}
S_{fix}&=&\int d^2\xi \left(\kappa v^a_+ \tilde \Pi_{+ab}v^b_--\frac{\kappa}{2} v^a_+ J_{-a}+\frac{\kappa}{2}J_{+a}v^a_-+{}_a j_{+}^M A_{MN} {}_a j_-^N\right)\nonumber \\
&+& \frac{\kappa}{2}\int d^2\xi  ( v_+^a\partial_- y_a-v^a_-\partial_+ y_a)\, .
\end{eqnarray}

On the equations of motion for $y_a$ we obtain that field strength is equal to zero
\begin{equation}
\partial_+ v_-^a-\partial_- v^a_+=0\, ,
\end{equation}
which solution is $v^a_\pm=\partial_\pm x^a$. In this way the action $S_{fix}$ turns to the initial action $S$. 

On the equations of motion for gauge fields $v^a_\pm$ we have
\begin{equation}\label{eq:y1}
\partial_+ y_a=2v^b_+ \tilde \Pi_{+ba}+J_{+a}\, ,
\end{equation}
\begin{equation}\label{eq:y2}
\partial_- y_a=-2\tilde \Pi_{+ab}v^b_-+J_{-a}\, .
\end{equation}
Substituting the expressions for $v_\pm^a$ 
\begin{equation}
v^a_+=\frac{1}{2}\left(\partial_+ y_b-J_{+b}\right)(\tilde \Pi_+^{-1})^{ba}\, ,
\end{equation}
\begin{equation}
v_-^a=-\frac{1}{2}(\tilde \Pi_+^{-1})^{ab}\left(\partial_-y_b-J_{-b}\right)\, ,
\end{equation}
into $S_{fix}$ we get T-dual action
\begin{eqnarray}\label{eq:Tdualdej}
{}_a S&=&\int d^2\xi \left[\frac{\kappa}{4}\partial_+ y_a (\tilde \Pi_+^{-1})^{ab} \partial_- y_b+\frac{1}{2}\partial_+ y_a (\tilde \Pi_+^{-1})^{ab}A_{bN}{}_a j_-^N-\frac{1}{2}{}_a j_+^M \bar A_{Ma}(\tilde \Pi_+^{-1})^{ab}\partial_- y_b\right.\nonumber \\
&+& \left. {}_a j_+^M\left(A_{MN}-\frac{1}{\kappa}\bar A_{Ma}(\tilde\Pi_+^{-1})^{ab}A_{bN}\right) {}_a j_-^N\right]\, ,
\end{eqnarray}
where we used the expressions for currents (\ref{eq:astrujap}) and (\ref{eq:astrujam}) and introduced the definitions
\begin{equation}
\bar A_{Ma}\equiv A_{MN} \bar P^N{}_a\, ,\quad A_{aM}\equiv P_a{}^N A_{NM}\, .
\end{equation}

\subsection{T-dual transformation laws}

In order to obtain relation between initial coordinates $x^a$ and corresponding T-dual ones $y_a$, we eliminate $v_\pm$ from the equations of motion 
for Lagrange multipliers $y_a$, $v^a_\pm =\partial_\pm x^a$, and other ones for gauge fields $v^a_\pm$ (\ref{eq:y1}) and (\ref{eq:y2})
\begin{equation}\label{eq:tduallaw}
\partial_\pm y_a\cong -2\tilde \Pi_{\mp ab} \partial_\pm x^b+J_{\pm a}\, .
\end{equation}
Using the expressions for currents ${}_a j_\pm^M=j_\pm^M+P^M{}_i\,\, \partial_\pm x^i$ given in Eq.(\ref{eq:strujejpmM}), we get the currents (\ref{eq:strujeJ}) in the form
\begin{equation}
J_{\pm a}=\bar J_{\pm a}-2\tilde \Pi_{\mp ai} \partial_\pm x^i\, ,
\end{equation}
where we introduced the notation
\begin{equation}
\bar J_{+a}=\frac{2}{\kappa}j_+^M A_{MN}\bar P^N{}_a\, ,\quad \bar J_{-a}=-\frac{2}{\kappa}P_a{}^M A_{MN}j^N_-\, ,
\end{equation}
\begin{equation}
\tilde \Pi_{+ia}\equiv \frac{1}{\kappa} P_i{}^M A_{MN}\bar P^N{}_a\, ,\quad \tilde\Pi_{+ai}=\frac{1}{\kappa}P_a{}^M A_{MN}\bar P^N{}_i\, .
\end{equation}

Now we can rewrite the transformation law (\ref{eq:tduallaw}) in the form
\begin{equation}\label{eq:tlawy}
\partial_\pm y_a\cong -2\tilde \Pi_{\mp ab}\partial_\pm x^b-2\tilde \Pi_{\mp ai}\partial_\pm x^i+\bar J_{\pm a}\, ,
\end{equation}
while the inverse one is
\begin{equation}\label{eq:tlawx}
\partial_\pm x^a\cong -2\kappa \tilde\theta_\pm^{ab}\tilde\Pi_{\mp bi} \partial_\pm x^i  -\kappa \tilde\theta^{ab}_{\pm}  ( \partial_\pm y_b  -  \bar J_{\pm b}  )   \, .
\end{equation}
Here we introduced the field $\tilde\theta_\pm^{ab}$ as
\begin{equation}
{\tilde \theta}^{ab}_\pm =  -\frac{2}{\kappa} ({\hat g}^{-1} {\tilde \Pi}_\pm {\tilde G}^{-1})^{ab}\, ,\quad {\hat g}_{ab} = ({\tilde G} -4 {\tilde B}  {\tilde G}^{-1} {\tilde B})_{ab}\, ,
\end{equation}
such that 
\begin{equation}\label{eq:tildetetapi}
\tilde \theta^{ab}_\pm\tilde \Pi_{\mp bc}=\frac{1}{2\kappa}\delta^a{}_c\, .
\end{equation}

Note that the form of the transformation laws is the same as in the case of constant background fields \cite{BNBSCBC}. 
But now all background fields depend on the undualized coordinates ($\theta^\alpha$, $\bar\theta^\alpha$, $x^i$).

Let us find relation between complete T-dual coordinates ${}_a X_{\hat \mu} = \{y_a , \, x^i \}$ and initial ones $x^\mu$. Together with T-dual transformation laws (\ref{eq:tlawy}), which relate $y_a$
 with $x^\mu$, we can add simply $\partial_\pm x^i=\partial_\pm x^i$ and rewrite both relations in the form 
\begin{equation}\label{eq:tdm}
\partial_+ ({}_a X)_{\hat \mu} = ({\bar Q}^{-1 T})_{{\hat \mu} \nu}  \partial_+ x^\nu  + \bar J_{+ \hat \mu}\, ,\quad
\partial_- \, ({}_a X)_{\hat \mu} = ( Q^{-1 T})_{{\hat \mu} \nu}  \partial_- x^\nu  +\bar J_{- \hat \mu}\, .
\end{equation}
The matrices
\begin{equation}\label{eq:qbarq}
 Q^{\hat \mu \nu}  =
 \left (
\begin{array}{cc}
\kappa {\tilde \theta}^{ab}_+   &    0        \\
  -2 \kappa \tilde\Pi_{- ic}  {\tilde \theta}^{cb}_+        &  \delta^i_j
\end{array}\right )  \, , \qquad
\bar Q^{\hat \mu \nu}  =
 \left (
\begin{array}{cc}
\kappa {\tilde \theta}^{ab}_-   &    0        \\
  -2 \kappa \tilde\Pi_{+ ic}  {\tilde \theta}^{cb}_-        &  \delta^i_j
\end{array}\right ) \, ,
\end{equation}
and theirs inverse
\begin{equation}\label{eq:qbarqi}
 Q^{-1}_{\mu \hat \nu}  =
 \left (
\begin{array}{cc}
2 \tilde\Pi_{- ab}  &    0        \\
 2 \tilde\Pi_{- ib}       &  \delta^j_i
\end{array}\right )  \, , \qquad
\bar Q^{-1}_{\mu \hat \nu}  =
 \left (
\begin{array}{cc}
2 \tilde \Pi_{+ ab}  &    0        \\
 2 \tilde\Pi_{+ ib}       &  \delta^j_i
\end{array}\right ) \,  ,
\end{equation}
perform T-dualization for vector indices. The currents are defined as
\begin{equation}
\bar J_{\pm \hat\mu}=\left ( \begin{array}{c}
\bar J_{\pm a}        \\
 0
\end{array}\right )\, .
\end{equation}

\subsection{Two sets of vielbeins and gamma matrices}

Different chiralities transform differently as in Refs.\cite{BPT,H,BNBSCBC}. Consequently, there are two types of T-dual vielbeins defined as
\begin{equation}\label{eq:tde}
{}_a e^{\underline{a} \hat \mu} = e^{\underline{a}}{}_\nu (Q^T)^{\nu \hat \mu} \, , \qquad
{}_a \bar e^{\underline{a} \hat \mu} = e^{\underline{a}}{}_\nu (\bar Q^T)^{\nu \hat \mu} \, ,
\end{equation}
producing the same T-dual metric ${}_a G^{\hat \mu \hat \nu}$, where hat indices are from T-dual picture. Two types of vielbeins produces two sets of gamma matrices in the
T-dual picture
\begin{equation}\label{eq:gamam1}
{}_a \Gamma_{\hat \mu} =  ({}_a e^{-1})_{ \hat \mu \underline{a}} \,   \Gamma^{\underline{a}} = ({}_a e^{-1} \Gamma)_{\hat \mu}  \, , \qquad
{}_a \bar \Gamma_{\hat \mu} =  ({}_a \bar e^{-1})_{ \hat \mu \underline{a}} \,   \Gamma^{\underline{a}}= ({}_a \bar e^{-1}  \Gamma)_{\hat \mu}   \, ,
\end{equation}
which are connected by local Lorentz transformation
\begin{equation}\label{eq:rgamam}
{}_a \bar \Gamma_{\hat \mu}  = {}_a \Omega^{-1} \, {}_a \Gamma_{\hat \mu} \; {}_a \Omega    \, .
\end{equation}
Here ${}_a \Omega$ is spinorial representation of the Lorentz transformation
\begin{equation}\label{eq:omega}
{}_a \Omega^{-1} \, \Gamma^{\underline{a}} \, \, {}_a \Omega = (\Lambda^{-1})^{\underline{a}}{}_{\underline{b}} \, \Gamma^{\underline{b}}    \, .
\end{equation}
The underlined indices are Lorentz ones (denoted by $\underline{a}, \underline{b}$). The matrix $\Lambda^{\underline{a}}{}_{\underline{b}}$ is a matrix of Lorentz transformation 
and it is given by the expression
\begin{equation}
\Lambda^{\underline{a}}{}_{\underline{b}}  = e^{\underline{a}}{}_\mu (Q^{-1} \bar Q)^{T \mu}{}_\nu  (e^{-1})^\nu{}_{\underline{b}}\, .
\end{equation}

In T-dual theory, as a consequence of two types of $\Gamma$ matrices, there are two types of T-dual supersymmetry invariant variables
\begin{equation}
{}_a d_\alpha={}_a \pi_\alpha-\frac{1}{2}({}_a\Gamma^{\hat\mu}\;{}_a\theta)_\alpha (\partial_+\;{}_a X_{\hat\mu}+\frac{1}{4}{}_a\theta {}_a\Gamma_{\hat\mu}\partial_+\;{}_a\theta)\, ,
\end{equation}
\begin{equation}
{}_a \bar d_\alpha={}_a \bar\pi_\alpha-\frac{1}{2}({}_a\bar\Gamma^{\hat\mu}\;{}_a\bar\theta)_\alpha (\partial_-\;{}_a X_{\hat\mu}+\frac{1}{4}{}_a\bar\theta {}_a\bar\Gamma_{\hat\mu}\partial_-\;{}_a\bar\theta)\, .
\end{equation}
In order to work with one set of gamma $\Gamma$ matrices we have to introduce proper variables. We can rewrite bar expression as
\begin{equation}
({}_a \Omega \;{}_a \bar d)_\alpha = ({}_a \Omega\;{}_a\bar\pi)_\alpha-\frac{1}{2}({}_a\Gamma^{\hat\mu}{}_a\Omega\;{}_a\bar\theta)_\alpha(\partial_-\;{}_a X_{\hat\mu}+\frac{1}{4}{}_a\bar\theta\;{}_a \Omega^{-1}{}_a\Gamma_{\hat\mu}\;{}_a\Omega\;\partial_- {}_a\bar\theta)\, .
\end{equation}
Let us preserve expressions for unbar variables, ${}_a\theta^\alpha=\theta^\alpha$ and ${}_a\pi_\alpha=\pi_\alpha$, and change bar variables
\begin{equation}\label{eq:bulet1}
{}^\bullet \bar\theta^\alpha\equiv {}_a \Omega^\alpha{}_\beta \;{}_a\bar\theta^\beta\, ,\quad {}^\bullet \bar\pi_\alpha\equiv {}_a \Omega_\alpha{}^\beta\;{}_a \bar\pi_\beta\, .
\end{equation}
Now the forms of transformation of the supersymmetric invariants are the same. In short, fermionic index without bar is unchanged, while bar fermionic index is multiplied by ${}_a \Omega$.

The further story, finding the spinorial representation of local 
Lorentz symmetry ${}_a\Omega$ connecting two kinds of vielbeins, is the same as in \cite{BPT,H,BNBSCBC} and we will not repeat it. We will just write the final expression for
matrix ${}_a\Omega$  in spinorial representation
\begin{equation}\label{eq:solom}
{}_a \Omega \,\, = {\sqrt{\prod_{i=1}^{d} G_{a_i a_i} }} \,\, {}_a  \Gamma \,  (i \, \Gamma^{11})^d \, ,
\end{equation}
\begin{equation}
\Gamma^{11}=(i)^{\frac{D(D-1)}{2}}\frac{1}{\prod_{\mu=0}^{D-1}G_{\mu\mu}} \varepsilon_{\mu_1\mu_2\dots \mu_D}\Gamma^{\mu_1}\Gamma^{\mu_2}\dots\Gamma^{\mu_D}\, .
\end{equation}
Matrix $\Gamma^{11}$ has normalization constant to satisfy the condition $(\Gamma^{11})^2=1$.
Also we have
\begin{equation}\label{eq:gama5}
{}_a  \Gamma \,  = (i)^{\frac{d(d-1)}{2}}  \prod_{i=1}^{d}  \Gamma^{a_i}  = (i)^{\frac{d(d-1)}{2}}\,  \Gamma^{a_1} \Gamma^{a_2} \cdots  \Gamma^{a_d} \,  ,
\end{equation}
so that
\begin{equation}\label{eq:gama52}
({}_a  \Gamma)^2 \,  =  \prod_{i=1}^{d} G^{a_i a_i} = \frac{1}{\prod_{i=1}^{d} G_{a_i a_i}} \, .
\end{equation}
This matrix is introduced as analogy of $\Gamma^{11}$ in subspace spanned by T-dualized directions $x^a$.
The letter $d$ denotes the number of T-dualized coordinates $x^a\;(a=1,2,\dots,d)$.

\section{Relations between initial and T-dual background fields}
\setcounter{equation}{0}

In this section we will find the most general form of the T-dual background fields in terms of the initial ones in the type II pure spinor superstring under simultaneous T-dualization. 
We will also discuss the form of the T-dual dilaton field obtained in the quantization procedure.

We expect that T-dual action (\ref{eq:Tdualdej}) has the form of the initial action (\ref{eq:vsg}) but expressed in terms of the T-dual variables and background fields
\begin{equation}\label{eq:dualnodej}
{}_a S=\int d^2\xi \;{}_a X^T_{\hat M} {}_a A^{\hat M\hat N} {}_a \bar X_{\hat N}\, ,
\end{equation}
where in analogy with (\ref{eq:strujejpmM}) we have
\begin{equation}
{}_a X_{\hat M}={}_a \hat P_{\hat M}{}^a \partial_+ y_a+{}^\star_a j_{+\hat M}\, ,\quad {}_a\bar X_{\hat M}={}_a\hat{\bar P}_{\hat M}{}^a \partial_- y_a+{}^\star_a \bar j_{-\hat M},\quad \left[\hat M=(\alpha,\hat \mu,\alpha,\hat\mu\hat\nu)\right]\, .
\end{equation}
Let us remind that index $\hat\mu$ means that index of T-dualized direction, $a$, goes up if it was down in initial theory and vice versa, while index $i$ keeps the position.

The decomposition of T-dual variables ${}_a X_{\hat M}$ and ${}_a \bar X_{\hat M}$ has similar form as for initial ones, $X^M$ and $\bar X^M$. 
We express the T-dual currents ${}^\star_a j_{+\hat M}$ and ${}^\star_a \bar j_{-\hat M}$ in terms of the initial ones ${}_a j_{\pm}^M$ 
as
\begin{equation}
{}^\star_a j_{+\hat M}=\omega_{\hat M N} {}_a j^N_+\, ,\quad {}^\star_a \bar j_{-\hat M}=\bar\omega_{\hat M N}{}_a j^N_-\, ,
\end{equation}
in order to make comparison of the actions (\ref{eq:dualnodej}) and (\ref{eq:Tdualdej}) which will produce the relations between T-dual and initial background fields. 
We did not write free field actions for pure spinors, $S_\lambda$ and $S_{\bar\lambda}$, because they carry fermionic indices while we T-dualize along some subset of 
bosonic indices. So, they are not changed in the process of T-dualization. 

Following the form of the initial theory we introduced for T-dual case
\begin{equation}
{}_a \hat P_{\hat M}{}^a = \left(
\begin{array}{c}
0 \\ \delta_a{}^b\\0\\ \frac{\kappa}{2}\tilde\theta^{ab}_-(\Gamma_b\theta)_\alpha\\0\, ,
\end{array}\right)\, ,\quad {}_a \hat{\bar P}_{\hat M}{}^a=\left(
\begin{array}{c}
0\\ \delta_a{}^b \\ 0\\ \frac{\kappa}{2}\tilde\theta^{ab}_+(\Gamma_b{}^\bullet\bar\theta)_\beta \Omega^\beta{}_\alpha\\ 0
\end{array}\right)\, .
\end{equation}
The matrices $\omega$ and $\bar\omega$ are of the form
\scriptsize{
\begin{equation}
\omega_{\hat M N}=\left(
\begin{array}{ccccc}
\delta^\alpha{}_\beta & 0 & 0 & 0 &0\\
\tilde\Pi_{-ai}(\theta\Gamma^i)_\alpha & 2\tilde \Pi_{-ab} & 0 & 0 & 0\\
-(\theta\Gamma^i)_\alpha & 0 & \delta^i{}_j & 0 & 0\\
\frac{\kappa}{2}\tilde\Pi_{+ic}\tilde\theta^{cb}_- (\Gamma_b \theta)_\alpha (\theta\Gamma^i)_\beta-\frac{1}{2}(\Gamma_i\theta)_\alpha(\theta\Gamma^i)^\beta & 0 & (\Gamma_i\theta)_\alpha-\kappa \tilde \Pi_{+ic}\tilde\theta^{cb}_- (\Gamma_b\theta)_\alpha & \delta_\alpha{}^\beta & 0\\
0 & 0 & 0 & 0 & (\bar Q^{-1})^T_{\hat \mu \rho} (\bar Q^{-1})^T_{\hat\nu\lambda}
\end{array}\right)\, ,
\end{equation}}
\normalsize
\tiny{\begin{equation}
\bar\omega_{\hat M N}=\left(
\begin{array}{ccccc}
\Omega^\alpha{}_\beta & 0 & 0 & 0 &0\\
\tilde\Pi_{+ai}({}^\bullet\bar\theta\Gamma^i)_\beta\; \Omega^\beta{}_\alpha & 2\tilde \Pi_{+ab} & 0 & 0 & 0\\
-({}^\bullet\bar\theta\Gamma^i)_\beta \Omega^\beta{}_\alpha & 0 & \delta^i{}_j & 0 & 0\\
(\frac{\kappa}{2}\tilde\Pi_{-ic}\tilde\theta^{cb}_+ (\Gamma_b {}^\bullet\bar\theta)_\gamma ({}^\bullet\bar\theta\Gamma^i)_\delta-\frac{1}{2}(\Gamma^i{}^\bullet\bar\theta)_\gamma ({}^\bullet\bar\theta\Gamma^i)_\delta) \Omega^\gamma{}_\alpha \Omega^\delta{}_\beta & 0 & [(\Gamma^i{}^\bullet\bar\theta)_\beta-\kappa \tilde \Pi_{-ic}\tilde\theta^{cb}_+ (\Gamma_b{}^\bullet\bar\theta)_\beta] \Omega^\beta{}_\alpha & {}_a\Omega_\alpha{}^\beta & 0\\
0 & 0 & 0 & 0 & (Q^{-1})^T_{\hat \mu \rho} (Q^{-1})^T_{\hat\nu\lambda}
\end{array}\right)\, .
\end{equation}}
\normalsize
We also will need inverse matrices, $\omega^{-1}$ and $\bar\omega^{-1}$,
\scriptsize{\begin{equation}
(\omega^{-1})^{M\hat N}=\left(
\begin{array}{ccccc}
\delta^\alpha{}_\beta & 0 & 0 & 0 & 0\\
-\frac{1}{2}(\tilde \Pi_-^{-1})^{ab}\tilde \Pi_{-bi}(\theta\Gamma^i)_\alpha & \frac{1}{2}(\tilde \Pi_-^{-1})^{ab} & 0 & 0 & 0\\
(\theta\Gamma^i)_\alpha & 0 & \delta^i{}_j & 0 & 0\\
\frac{\kappa}{2}\tilde\Pi_{+ic}\tilde\theta^{cb}_-(\Gamma_b\theta)_\alpha (\theta\Gamma^i)_\beta-(\Gamma_i\theta)_\alpha (\theta\Gamma^i)_\beta & 0 & \kappa \tilde\Pi_{+ic}\tilde\theta^{cb}_-(\Gamma_b\theta)_\alpha-(\Gamma_i\theta)_\alpha & \delta_\alpha{}^\beta & 0\\
0 & 0 & 0 & 0 & (\bar Q^T)^{\mu\hat\rho}(\bar Q^T)^{\nu\hat\lambda}
\end{array}\right)\, ,
\end{equation}}
\normalsize
\tiny{\begin{equation}
(\bar\omega^{-1})^{M\hat N}=\left(
\begin{array}{ccccc}
\Omega^\alpha{}_\beta & 0 & 0 & 0 & 0\\
-\frac{1}{2}(\tilde \Pi_+^{-1})^{ab}\tilde \Pi_{+bi}({}^\bullet\bar\theta\Gamma^i)_\alpha \Omega^\alpha{}_\beta & \frac{1}{2}(\tilde \Pi_+^{-1})^{ab} & 0 & 0 & 0\\
({}^\bullet\bar\theta\Gamma^i)_\alpha \Omega^\alpha{}_\beta & 0 & \delta^i{}_j & 0 & 0\\
\left[\frac{\kappa}{2}\tilde\Pi_{-ic}\tilde\theta^{cb}_+(\Gamma_b{}^\bullet\bar\theta)_\alpha ({}^\bullet\bar\theta\Gamma^i)_\beta-(\Gamma_i {}^\bullet\bar\theta)_\alpha ({}^\bullet\bar\theta\Gamma^i)_\beta\right]\Omega^\alpha{}_\gamma \Omega^\beta{}_\delta & 0 & [\kappa \tilde\Pi_{-ic}\tilde\theta^{cb}_+(\Gamma_b{}^\bullet\bar\theta)_\alpha-(\Gamma_i{}^\bullet\bar\theta)_\alpha]\Omega^\alpha{}_\beta & \Omega_\alpha{}^\beta & 0\\
0 & 0 & 0 & 0 & (Q^T)^{\mu\hat\rho}(Q^T)^{\nu\hat\lambda}
\end{array}\right)\, .
\end{equation}}
\normalsize

During the calculation of above matrices we used the expressions for T-dual gamma matrices with upper and lower indices
\begin{equation}
{}_a\Gamma_{\hat\mu}=-(Q^{-1})^T_{\hat\mu\nu}\Gamma^\nu=\left(
\begin{array}{c}
{}_a\Gamma_a\\ {}_a\Gamma^i
\end{array}\right)
=\left(
\begin{array}{c}
2\tilde\Pi_{+ab}\Gamma^b+2\tilde \Pi_{+ai}\Gamma^i\\
-\Gamma^i
\end{array}\right)\, ,
\end{equation}
\begin{equation}
{}_a\Gamma^{\hat\mu}=-Q^{\hat\mu\nu}\Gamma_\nu=\left(
\begin{array}{c}
{}_a\Gamma^a\\ {}_a\Gamma_i
\end{array}\right)
=\left(
\begin{array}{c}
-\kappa\tilde\theta^{ab}_+ \Gamma_b\\
-\Gamma_i+2\kappa \tilde\Pi_{-ic}\tilde\theta^{cb}_+ \Gamma_b
\end{array}\right)\, ,
\end{equation}
\begin{equation}
{}_a \bar\Gamma_{\hat\mu}=-(\bar Q^{-1})^T_{\hat\mu \nu}\Gamma^\nu=\left(
\begin{array}{c}
{}_a \bar\Gamma_{a}\\\bar\Gamma^i
\end{array}\right)=
\left(
\begin{array}{c}
2\tilde \Pi_{-ab}\Gamma^b+2\tilde\Pi_{-ai}\Gamma^i\\
-\Gamma^i
\end{array}\right)\, ,
\end{equation}
\begin{equation}
{}_a\bar \Gamma^{\hat\mu}=-\bar Q^{\hat\mu \nu}\Gamma_\nu=\left(
\begin{array}{c}
{}_a \bar\Gamma^a\\{}_a \bar\Gamma_i 
\end{array}\right)=
\left(
\begin{array}{c}
-\kappa \tilde\theta^{ab}_{-}\Gamma_b\\
-\Gamma_i+2\kappa \tilde \Pi_{+ic}\tilde\theta^{cb}_- \Gamma_b
\end{array}\right)\, .
\end{equation}

The explicit form of the action given in (\ref{eq:dualnodej}) is
\begin{eqnarray}
{}_a S&=&\int d^2\xi \left( \partial_+ y_a \hat P^{T\; a}{}_{\hat M} \;{}_a A^{\hat M\hat N} \hat{\bar P}_{\hat N}{}^b \partial_- y_b+\partial _+ y_a \hat P^{T\;a}{}_{\hat M}\;{}_a A^{\hat M\hat N}\bar\omega_{\hat N P}\;{}_a j^P_-\right.\nonumber\\
&+&\left. {}_a j^N_+ \omega^T_{N\hat M}\; {}_a A^{\hat M\hat N}\hat{\bar P}_{\hat N}{}^b \partial_- y_b+{}_a j_+^M \omega^T_{M\hat P}\;{}_a A^{\hat P\hat Q}\bar\omega_{\hat Q N}{}_a j_-^N\right)\, .
\end{eqnarray}
Comparing this action with that from Eq.(\ref{eq:Tdualdej}) we obtain the T-dual fields in terms of initial ones
\begin{equation}\label{eq:tdl1}
\hat P^{T\; a}{}_{\hat M} \; {}_a A^{\hat M\hat N}\hat{\bar P}_{\hat N}{}^b=\frac{\kappa}{4}(\tilde \Pi^{-1}_+)^{ab}\Longrightarrow {}_a\tilde \Pi_+^{ab}=\frac{1}{4}(\tilde\Pi_+^{-1})^{ab}\, ,
\end{equation}
\begin{equation}\label{eq:tdl2}
{}_a A^{a\hat M}=\frac{1}{2}(\tilde\Pi_+^{-1})^{ab} A_{b P}(\bar \omega^{-1})^{P\hat M}\, ,
\end{equation}
\begin{equation}\label{eq:tdl3}
{}_a \bar A^{\hat M a}= -\frac{1}{2}{(\omega^T)^{-1}}^{\hat M N}\bar A_{Nb}(\tilde \Pi^{-1}_+)^{ba}
\end{equation}
\begin{equation}\label{eq:tdl4}
{}_a A^{\hat M\hat N}={(\omega^T)^{-1}}^{\hat M P}\left(A_{PQ}-\frac{1}{\kappa}\bar A_{P a}(\tilde\Pi_+^{-1})^{ab}A_{bQ}\right)(\bar\omega^{-1})^{Q\hat N}\, ,
\end{equation}
where
\begin{equation}
{}_a A^{a\hat N}\equiv \hat P^{T\;a}{}_{\hat M}\; {}_a A^{\hat M\hat N}\, ,\quad {}_a \bar A^{\hat M a}\equiv {}_a A^{\hat M\hat N}\hat{\bar P}_{\hat N}{}^a\, .
\end{equation}
The next step is to express components of the T-dual fields in terms of the components of the initial background fields.
Also in order to find transformation law for physical superfields components in $A_{MN}$ we need the explicit expressions
\begin{equation}\label{eq:vek1}
{}_a A^{a\hat N}=\left(
\begin{array}{c}
{}_a A^a{}_\alpha+\frac{\kappa}{2}\tilde\theta^{ab}_- (\Gamma_b\theta)_\beta \;{}_a E^\beta{}_\alpha\\
{}_a A^{ab}+\frac{\kappa}{2}\tilde\theta^{ab}_- (\Gamma_b\theta)_\beta \;{}_a E^{\beta b}\\
{}_a A^a{}_i+\frac{\kappa}{2}\tilde\theta^{ab}_- (\Gamma_b\theta)_\beta \;{}_a E^\beta{}_i\\
{}_a \bar E^{a\beta}+\frac{\kappa}{2}\tilde\theta^{ab}_- (\Gamma_b\theta)_\alpha \;{}_a P^{\alpha\beta}\\
{}_a \Omega^{a,\hat \mu\hat\nu}+\frac{\kappa}{2}\tilde\theta^{ab}_- (\Gamma_b\theta)_\alpha \;{}_a C^{\alpha,\hat\mu\hat\nu}
\end{array}\right)^T\, ,
\end{equation}
\begin{equation}
{}_a \bar A^{\hat M a}=\left(
\begin{array}{c}
{}_a A_\alpha{}^a+\frac{\kappa}{2}{}_a E_\alpha{}^\beta \tilde \theta^{ab}_+ (\Gamma_b {}^\bullet \bar\theta)_\gamma \Omega^\gamma{}_\beta\\
{}_a A^{ab}+\frac{\kappa}{2}{}_a \bar E^{a\beta}\tilde \theta^{bc}_+ (\Gamma_c {}^\bullet \bar\theta)_\gamma \Omega^\gamma{}_\beta\\
{}_a A_i{}^a+\frac{\kappa}{2}{}_a \bar E_i{}^\beta \tilde \theta^{ab}_+ (\Gamma_b {}^\bullet \bar\theta)_\gamma \Omega^\gamma{}_\beta\\
{}_a E^{\alpha a}+\frac{\kappa}{2} {}_a P^{\alpha\beta}\tilde \theta^{ab}_+ (\Gamma_b {}^\bullet \bar\theta)_\gamma \Omega^\gamma{}_\beta\\
{}_a \Omega^{\hat\mu\hat\nu,a}+\frac{\kappa}{2}{}_a\bar C^{\hat\mu\hat\nu,\beta}\tilde \theta^{ab}_+ (\Gamma_b {}^\bullet \bar\theta)_\gamma \Omega^\gamma{}_\beta
\end{array}\right)\, ,
\end{equation}
\begin{equation}
A_{aM}=\left(
\begin{array}{c}
A_{a\alpha}-\frac{1}{2}(\Gamma_a\theta)_\beta E^\beta{}_\alpha\\
A_{ab}-\frac{1}{2}(\Gamma_a\theta)_\alpha E^\alpha{}_b\\
A_{ai}-\frac{1}{2}(\Gamma_a\theta)_\alpha E^\alpha{}_i\\
\bar E_a{}^\alpha-\frac{1}{2}(\Gamma_a\theta)_\beta P^{\beta\alpha}\\
\Omega_{a,\mu\nu}-\frac{1}{2}(\Gamma_a\theta)_\alpha C^\alpha{}_{\mu\nu}
\end{array}\right)^T\, ,
\end{equation}
\begin{equation}\label{eq:vek2}
\bar A_{Ma}=\left(
\begin{array}{c}
A_{\alpha a}-\frac{1}{2}E_\alpha{}^\beta (\Gamma_a\bar\theta)_\beta\\
A_{ab}-\frac{1}{2}\bar E_a{}^\beta (\Gamma_b\bar\theta)_\beta\\
A_{ia}-\frac{1}{2}\bar E_i{}^\beta(\Gamma_a\bar\theta)_\beta\\
E^\alpha{}_a-\frac{1}{2}P^{\alpha\beta}(\Gamma_a\bar\theta)_\beta\\
\Omega_{\mu\nu,a}-\frac{1}{2}\bar C_{\mu\nu}{}^\beta (\Gamma_a\bar\theta)_\beta
\end{array}\right)\, .
\end{equation}

In order to describe  dilaton field $\Phi$ in the standard formulation one should  add Fradkin-Tseytlin term as in \cite{BPT}
\begin{equation}\label{eq:FTa}
S_\Phi = \int d^2 \xi \sqrt{-g} R^{(2)} \Phi  \, ,
\end{equation}
to the initial action. Here $R^{(2)}$ is scalar curvature of the world sheet.  
It is well known that dilaton field transformation under T-dualization is considered within path integral formalism \cite{B,H,BPT,BNBSCBC,GR}. 
For a constant background the Gaussian path integral produces the expression $(\det \tilde\Pi_{ + a b})^{-1}$.

In this article we T-dualize just along a subset of coordinates $x^a$ and assume that all background fields are independent of $x^a$. Consequently, gaussian integration over gauge fields $v^a_\pm$ 
in path integral produces the same result as in the so called constant background case \cite{BNBSCBC} we get the form of the T-dual dilaton field.
\begin{equation}\label{eq:dsh}
 {}_a \Phi(x^i,\theta^\alpha,\bar\theta^\alpha)  = \Phi(x^i,\theta^\alpha,\bar\theta^\alpha) - \ln \det ( 2 \tilde\Pi_{+ ab}) \, .
\end{equation}
Using the expression for $\tilde \Pi_{+ab}$ (\ref{eq:tildepiab}) in the form
\begin{equation}
\tilde \Pi_{+ab}=\Pi_{+ab}-\Delta_{ab}\, ,
\end{equation}
where $\Delta_{ab}$ is defined as
\begin{equation}\label{eq:Deltaab}
\Delta_{ab}=\frac{1}{2\kappa}\bar E_a{}^\alpha (\Gamma_b \bar\theta)_\alpha+\frac{1}{2\kappa}(\Gamma_a \theta)_\alpha E^\alpha{}_b-\frac{1}{4\kappa}(\Gamma_a \theta)_\alpha P^{\alpha\beta}(\Gamma_b \bar\theta)_\beta\, ,
\end{equation}
we get
\begin{equation}\label{eq:Tdilaton}
{}_a \Phi(x^i,\theta^\alpha,\bar\theta^\alpha)  = \Phi(x^i,\theta^\alpha,\bar\theta^\alpha) - \ln \det ( 2\Pi_{+ ac}) -\ln \det(\mathbbm{1}-\Pi_+^{-1}\Delta)^c{}_b\, .
\end{equation}
Quantity $\Delta_{ab}$  measures the difference between general case considered in this article and the case considered in Ref.\cite{BNBSCBC}.  
Note that the expressions for background fields in Ref.\cite{BNBSCBC} have been obtained from general ones in Ref.\cite{verteks} after elimination of all $\theta^\alpha$ and $\bar\theta^\alpha$ dependent terms in the action. 
In that manner in Ref.\cite{BNBSCBC} we omitted all terms in  $\Delta_{ab}$. From Eq.(4.26) we see that ignoring $\theta^\alpha$ and $\bar\theta^\alpha$ dependent terms in \cite{BNBSCBC} prevents us to get complete solution for the T-dual dilaton.

\section{The components of the T-dual matrix ${}_a A^{\hat M \hat N}$}
\setcounter{equation}{0}

In this section we will write explicit expressions for T-dual superfields and, putting that background fields are constant, compare the results with already known constant background case.

\subsection{The physical superfields and comparison with constant background case}

In order to find T-dual field ${}_a A^{ab}$ we take into consideration some particular component of the equations (\ref{eq:tdl2})-(\ref{eq:tdl4}).

{\bf The second component of Eq.(\ref{eq:tdl2}) ($\hat M\to b$)}. 
It produces the equation
\begin{equation}\label{eq:comp1}
{}_a A^{ab}+\frac{\kappa}{2}\tilde\theta^{ab}_-(\Gamma_b\theta)_\alpha \; {}_a E^{\alpha b}=\frac{1}{2}(\tilde \Pi_+^{-1})^{ac}\left[A_{cd}-\frac{1}{2}(\Gamma_c\theta)^\alpha E^\alpha{}_d\right]\cdot \frac{1}{2}(\tilde\Pi_+^{-1})^{db}\, .
\end{equation}
We treat left-hand side and right-hand side of this equation as an expansion in powers of $\theta^\alpha$. Equating appropriate coefficients we obtain the T-dual fields ${}_a A^{ab}$ and ${}_a E^{\alpha a}$
\begin{equation}
{}_a A^{ab}=\frac{1}{4}(\tilde \Pi_+^{-1})^{ac} A_{cd} (\tilde \Pi_+^{-1})^{db}\, ,\quad {}_a E^{\alpha a}=\frac{1}{2}(\tilde \Pi_-^{-1})^{ab}E_b{}^\alpha\, . \quad \left(\tilde \Pi_-=-\Pi_+^T\right )
\end{equation}
Using the redefinitions $A_{ab}=\kappa\Pi_{+ab}$ and ${}_a A^{ab}=\kappa \;{}_a \Pi_{+}^{ab}$ as well as the relation $\tilde\theta^{ab}_-\tilde \Pi_{+bc}=\frac{1}{2\kappa}\delta^a{}_c$, we get
\begin{equation}\label{eq:piab}
{}_a \Pi^{ab}_+=\frac{1}{4}(\tilde \Pi_+^{-1})^{ac}\Pi_{+cd} (\tilde \Pi_+^{-1})^{db} \, ,
\end{equation}
\begin{equation}\label{eq:eab}
{}_a E^{\alpha a}=\kappa \tilde\theta^{ab}_+ E^\alpha{}_b\, .
\end{equation}
In the case of the constant background fields ($\Delta_{ab}=0$), the relation (\ref{eq:piab}) transforms into
\begin{equation}\label{eq:poredj0}
{}_a \Pi_+^{ab}=\frac{1}{4}(\Pi_+^{-1})^{ab}=\frac{\kappa}{2}\hat\theta^{ab}_-\, ,
\end{equation}
because in that case $\tilde \Pi_{+ab}=\Pi_{+ab}$ and $\tilde\theta^{ab}_\pm\to \hat\theta^{ab}_\pm$.

The equation (\ref{eq:eab}) in the limit of the constant background fields is in accordance with 
appropriate result obtained in \cite{BNBSCBC}. Here we have in mind that field $\Psi^\alpha_\mu$ is zero order term in the expansion of $E^\alpha_\mu$ \cite{verteks}, which produces
\begin{equation}\label{eq:poredj1}
{}_a\Psi^{\alpha a}=\kappa \hat \theta^{ab}_+ \Psi^\alpha_b\, .
\end{equation}

{\bf The second component of Eq.(\ref{eq:tdl3}) ($\hat M\to a$)}. It produces
\begin{equation}\label{eq:comp2}
{}_a A^{ab}+\frac{\kappa}{2}{}_a \bar E^{a\beta}\tilde\theta^{bc}_+(\Gamma_c {}^\bullet \bar\theta)_\gamma \Omega^\gamma{}_\beta=-\frac{1}{2}(-\frac{1}{2})(\tilde \Pi_+^{-1})^{ac}
\left[A_{cd}-\frac{1}{2}\bar E_c{}^\beta (\Gamma_d\bar\theta)_\beta\right](\tilde \Pi_+^{-1})^{db}\, ,
\end{equation}
we can get again the expression for ${}_a A^{ab}$, but we additionally have
\begin{equation}
{}_a \bar E^{a\alpha}=\kappa \tilde\theta^{ab}_- \bar E_b{}^\beta \Omega_\beta{}^\alpha\, .
\end{equation}
In the constant background case $\bar E_a{}^\alpha\to\bar \Psi_a{}^\alpha$. Consequently, here we also have good constant background limit
\begin{equation}\label{eq:poredj2}
{}_a\bar\Psi^{\alpha a}=\kappa \hat\theta^{ab}_- \bar\Psi_b{}^\beta \Omega_\beta{}^\alpha\, .
\end{equation}

It is useful to observe the fact that expressions for T-dual fields ${}_a \bar E^{a\alpha}$ and ${}_a E^{a\alpha}$ can be obtained analyzing
{\bf the fourth components of the equations (\ref{eq:tdl2}) and (\ref{eq:tdl3})}, respectively. Also 
note that $\hat\theta^{ab}_{\pm}$ appearing in (\ref{eq:poredj1}) and (\ref{eq:poredj2}) is constant tensor defined as inverse of the constant tensor $\Pi_{+ab}$ \cite{BNBSCBC}.

{\bf The third component of Eq.(\ref{eq:tdl2}) ($\hat M\to i$)}.
Let us consider the equation which follows from the third component of the Eq.(\ref{eq:tdl2}) ($\hat M\to i$)
\begin{eqnarray}\label{eq:comp3}
&{}& {}_a A^a{}_i+\frac{\kappa}{2}\tilde\theta^{ab}_- (\Gamma_b \theta)_\beta {}_a E^\beta{}_i=\frac{1}{2}(\tilde \Pi_+^{-1})^{ab}\left(A_{bj}-\frac{1}{2}(\Gamma_b\theta)_\beta E^\beta{}_i\right) \nonumber\\
&+& \frac{1}{2}(\tilde \Pi_+^{-1})^{ab}\left(\bar E_b{}^\alpha-\frac{1}{2}(\Gamma_b\theta)_\beta P^{\beta\alpha}\right)\left(\kappa\tilde\Pi_{-ic}\tilde\theta^{cb}_+ (\Gamma_b{}^\bullet \bar\theta)_\gamma-(\Gamma_i {}^\bullet\bar\theta)_\gamma\right)\Omega^\gamma{}_\alpha\, .
\end{eqnarray}
Extracting the zero components in the expansion we get
\begin{equation}
{}_a A^a{}_i=\frac{1}{2}(\tilde\Pi_+^{-1})^{ab}A_{bi}\, .
\end{equation}
In order to make comparison with the constant background case easier, we introduce the following notation
\begin{equation}
{}_a A^a{}_i=\kappa \;{}_a\Pi^a_{+i}\, ,\quad A_{ai}=\kappa \Pi_{+ai}\, ,
\end{equation}
and, using the relation (\ref{eq:tildetetapi}), we obtain
\begin{equation}\label{eq:poredj3}
{}_a \Pi^a_{+i}=\kappa \tilde\theta^{ab}_-\Pi_{+bi}\, .
\end{equation}
Treating {\bf the third component of the Eq.(\ref{eq:tdl3}) ($\hat M\to i$)}
\begin{eqnarray}\label{eq:comp4}
&{}& {}_a A_i{}^a+\frac{\kappa}{2}{}_a \bar E_i{}^\alpha \tilde\theta^{ab}_+ (\Gamma_b {}^\bullet\bar\theta)_\beta \Omega^\beta{}_\alpha=-\frac{1}{2}\left(A_{ib}-\frac{1}{2}\bar E_i{}^\beta(\Gamma_b\bar\theta)_\beta\right)\nonumber\\
&-&\frac{1}{2}\left(\kappa \tilde\Pi_{+ic}\tilde\theta^{cb}_- (\Gamma_b\theta)_\alpha-(\Gamma_i\theta)_\alpha\right)\left(E^\alpha{}_d-\frac{1}{2}P^{\alpha\beta}(\Gamma_d\bar\theta)_\beta\right)(\tilde \Pi_+^{-1})^{da}\, ,
\end{eqnarray}
in the same way as in the previous case, we have
\begin{equation}\label{eq:poredj4}
{}_a \Pi_{+i}{}^a=-\kappa \Pi_{+ib}\tilde\theta^{ba}_{-}\, .
\end{equation}
The last two expressions, (\ref{eq:poredj3}) and (\ref{eq:poredj4}), in the limit of the constant background fields are in full correspondence with the result obtained in the constant background case \cite{BNBSCBC}.
\vskip0.5cm

{\bf The ($\hat M\to i\, ,\hat N\to j$) component of Eq.(\ref{eq:tdl4}).}
Considering appropriate component in the Eq.(\ref{eq:tdl4}) ($\hat M\to i\, ,\hat N\to j$), we obtain
\begin{equation}\label{eq:comp5}
{}_a A_{ij}=A_{ij}-\frac{1}{\kappa}\left[A_{ia}-\frac{1}{2}\bar E_i{}^\alpha(\Gamma_a\bar\theta)_\alpha\right](\tilde \Pi_+^{-1})^{ab}\left[A_{bj}-\frac{1}{2}(\Gamma_b\theta)_\delta E^\delta{}_j\right]\, .
\end{equation}
Using redefinitions, $A_{ij}=\kappa \Pi_{+ij}$ and ${}_a A_{ij}=\kappa\;{}_a \Pi_{+ij}$, this relation can be rewritten in the form
\begin{equation}
{}_a \Pi_{+ij}=\Pi_{+ij}-\frac{1}{\kappa^2}\left[\kappa \Pi_{+ia}-\frac{1}{2}\bar E_i{}^\alpha(\Gamma_a\bar\theta)_\alpha\right](\tilde \Pi_+^{-1})^{ab}\left[\kappa \Pi_{+bj}-\frac{1}{2}(\Gamma_b\theta)_\delta E^\delta{}_j\right]\, .
\end{equation}
In the constant background case explicit $\theta^\alpha$ and $\bar\theta^\alpha$ dependence disappears and $\tilde \Pi_{+ab}=\Pi_{+ab}$. Consequently, we get
\begin{equation}\label{eq:poredj5}
{}_a\Pi_{+ij}=\Pi_{+ij}-2\kappa \Pi_{+ia} \hat\theta^{ab}_- \Pi_{+bj}\, ,
\end{equation}
which is exactly the relation obtained in the constant background case \cite{BNBSCBC}.
\vskip0.5cm

{\bf Eq.(\ref{eq:tdl4}) where $(\hat M\to i, \hat N\to \alpha)$, $(\hat M\to \alpha, \hat N\to i)$ and $(\hat M\to\alpha, \hat N\to \beta)$.}
Also we read in this case, respectively,  
\begin{equation}\label{eq:comp6}
{}_a \bar E_i{}^\alpha=\{E_i{}^\gamma-\frac{1}{\kappa} \left[A_{ia}-\frac{1}{2}\bar E_i{}^\beta(\Gamma_a \bar\theta)_\beta\right](\tilde \Pi_+^{-1})^{ab}\left[\bar E_b{}^\alpha-\frac{1}{2}(\Gamma_b\theta)_\delta P^{\delta\gamma}\right]\}\Omega_\gamma{}^\alpha\, ,
\end{equation}
\begin{equation}\label{eq:comp7}
{}_a E^\alpha{}_i=E^\alpha{}_i-\frac{1}{\kappa}\left[E^\alpha{}_a-\frac{1}{2}P^{\alpha\beta}(\Gamma_a\bar\theta)_\beta\right](\tilde \Pi_+^{-1})^{ab}\left[A_{bi}-\frac{1}{2}(\Gamma_b\theta)_\gamma E^\gamma{}_i\right]\, ,
\end{equation}
\begin{equation}\label{eq:comp8}
{}_a P^{\alpha\beta}=\{P^{\alpha\gamma}-\frac{1}{\kappa}\left[E^\alpha{}_a-\frac{1}{2}P^{\alpha\delta}(\Gamma_a\bar\theta)_\delta\right](\tilde \Pi_+^{-1})^{ab}\left[\bar E_b{}^\gamma-\frac{1}{2}(\Gamma_b\theta)_\epsilon P^{\epsilon\gamma}\right]\}\Omega_\gamma{}^\beta\, .
\end{equation}
In the constant background limit which effectively means that we put $\theta^\alpha=\bar\theta^\alpha=0$, $A_{ia}=\kappa\Pi_{+ia}$, $A_{bi}=\kappa\Pi_{+bi}$, $(\tilde\Pi_{+}^{-1})^{ab}=2\kappa \hat\theta^{ab}_-$ and 
$P^{\alpha\beta}=\frac{1}{2\kappa}e^{\frac{\Phi}{2}}F^{\alpha\beta}$, we obtain relations from \cite{BNBSCBC}
\begin{equation}\label{eq:poredj6}
{}_a \bar \Psi_i{}^\alpha=\left[\bar \Psi_i{}^\beta-2\kappa\Pi_{+ia}\hat\theta^{ab}_- \bar \Psi_b{}^\beta\right]\Omega_\beta{}^\alpha\, ,
\end{equation}
\begin{equation}\label{eq:poredj7}
{}_a \Psi^\alpha{}_i=\Psi^\alpha{}_i-2\kappa \Psi^\alpha{}_a \hat\theta^{ab}_- \Pi_{+bi}\, ,
\end{equation}
\begin{equation}\label{eq:poredj8}
e^{\frac{{}_a \Phi}{2}}{}_a F^{\alpha\beta}=\left[e^{\frac{\Phi}{2}}F^{\alpha\gamma}-4\kappa \Psi^\alpha{}_a \hat\theta^{ab}_- \bar \Psi_b{}^\gamma\right]\Omega_\gamma{}^\beta\, .
\end{equation}

Our compact result of the general case (\ref{eq:tdl2})-(\ref{eq:tdl4}) in components has a form (\ref{eq:comp1}), (\ref{eq:comp2}), (\ref{eq:comp3}), (\ref{eq:comp4}), (\ref{eq:comp5}) and (\ref{eq:comp6})-(\ref{eq:comp8}).
It gives the right limit for the constant background fields (\ref{eq:poredj0}), (\ref{eq:poredj1}), (\ref{eq:poredj2}), (\ref{eq:poredj3}),
(\ref{eq:poredj4}), (\ref{eq:poredj5}) and (\ref{eq:poredj6})-(\ref{eq:poredj8}).

\subsection{T-dual auxiliary background fields and field strengths}

The part of the main result, beside full expressions for the T-dual physical superfields, are the expressions for T-dual auxiliary superfields (the first column and the first row in
matrix ${}_a A^{\hat M\hat N}$) and expressions for T-dual field strengths (the last column and the last row in matrix ${}_a A^{\hat M\hat N}$). These background fields are absent in the already considered constant background
case \cite{BNBSCBC}, because imposed assumptions eliminated them from the theory (detailed argumentation is in \cite{BNBSCBC} and \cite{verteks}).

In order to read all mentioned background fields we use the appropriate components of (\ref{eq:tdl2})-(\ref{eq:tdl4}). Fixing the indices $\hat M$ and $\hat N$, we get the equations which we treat, as in the previous cases,
as expansions in powers of $\theta^\alpha$ and $\bar\theta^\alpha$. Equating the appropriate coefficients in the expansions, we read the form of T-dual background fields. Because there are many expressions 
and some
of them are long, we write just a few of them.
For example, we give the form of the background field T-dual to the $A_{a\alpha}$ and field strength T-dual to the $\Omega_{a,\mu\nu}$. Using the relations (\ref{eq:tdl2}) and (\ref{eq:vek1})-(\ref{eq:vek2}), after straightforward calculation, we get
\begin{equation}\label{eq:uvod1}
(\hat M\to \alpha)\quad {}_a A^a{}_\alpha=\kappa \tilde\theta^{ab}_- A_{b\beta}\;{}_a\Omega^\beta{}_\alpha\, ,
\end{equation}
\begin{equation}\label{eq:uvod2}
(\hat M\to \hat \mu\hat\nu)\quad {}_a\Omega^{a,\hat \mu\hat \nu}=\kappa \tilde\theta^{ab}_- \Omega_{b,\rho\lambda}(Q^T)^{\rho\hat\mu}(Q^T)^{\lambda\hat\nu}\, ,
\end{equation}
while considering the component of Eq.(\ref{eq:tdl4}) where $\hat M\to \hat\mu\hat\nu$ and $\hat N\to \hat\lambda\hat\rho$, we obtain
\begin{eqnarray}
{}_a S^{\hat\mu\hat\nu,\hat\lambda\hat\rho}&=&\bar Q^{\hat\mu \mu}\bar Q^{\hat\nu\nu} \left[ S_{\mu\nu,\lambda\rho}-\left(\Omega_{\mu\nu,a}-\frac{1}{2}\bar C_{\mu\nu}{}^\alpha(\Gamma_a \bar\theta)_\alpha\right)\right.\nonumber\\
&{}&\cdot \left. \tilde \theta^{ab}_-\left(\Omega_{b,\lambda\rho}-\frac{1}{2}(\Gamma_b\theta)_\beta C^\beta{}_{\lambda\rho}\right)\right](Q^T)^{\lambda\hat\lambda}(Q^T)^{\rho\hat\rho}\, .
\end{eqnarray}

\section{Concluding remarks}
\setcounter{equation}{0}

In this paper we have investigated simultaneous T-dualization of the pure spinor type II superstring described by the action of Ref.\cite{verteks}. We assumed that background fields do not depend on 
the coordinates along which we make T-dualization. Our goal was to find the form of the T-dual background fields, especially  T-dual auxiliary fields and field strengths which are not present in the constant 
background case. 
In relation to the articles \cite{H,BPT}, where single direction T-dualization is performed, here we demonstrated simultaneous T-dualization along some
subset, $x^a\;(a=1,2,\dots d)$,  of space-time directions. Also following Refs.\cite{H,BPT,BNBSCBC}, we found the form of the spinorial representation of local Lorentz 
transformation ${}_a\Omega$ occurring in the
T-dual picture.

The action we used in this article is type II superstring action in pure spinor formulation of Ref.\cite{verteks}. It is derived using nilpotency and (anti)holomorphicity conditions as
an expansion in powers of $\theta^\alpha$ and $\bar\theta^\alpha$. In Ref.\cite{BNBSCBC} we considered constant background version of this action obtained under certain assumptions - 
background fields are independent of all $x^\mu$ coordinates and we take just first components in the expansions of background fields. In this way we lost information about the form
of the T-dual auxiliary background fields and field strengths, as well as the complete form of T-dual physical superfields. 

It is difficult to work with the expanded form of action (\ref{eq:VSG}) because it has large number of terms. We used condensed form of the action (\ref{eq:vsg}) and extracted in
variables $X^M$ and $\bar X^M$ 
terms containing derivatives of the directions 
along we T-dualize, $\partial_\pm x^a$. The rest part of these variables is denoted as current ${}_a j_{\pm}^M$.
We inserted that expression into action and made T-dualization along $x^a$ direction. On the equation of motion for gauge fields $v_\pm^a$ we obtained T-dual action expressed in terms of T-dual coordinates $y_a$ and currents ${}_a j_\pm ^M$. Under T-dualization the form of the action is preserved and consequently, expressing the T-dual action in terms of the T-dual variables and fields, we finally got all T-dual background fields in the 
considered general case. In order to compare them with the constant background case of Ref.\cite{BNBSCBC} we explicitly wrote the expressions for physical superfields. In the limit of constant background fields obtained expressions
turn into the expressions of Ref.\cite{BNBSCBC}.

Combining the equations of motion for Lagrange multipliers $y_a$ and for gauge fields $v_\pm^a$ we obtain T-dual transformation laws (\ref{eq:tdm}) in most general case of type II pure spinor superstring.
Let us stress that we consider the general case and that all background fields now depend on the undualized directions $x^i$, $\theta^\alpha$ and $\bar\theta^\alpha$. Because
two chiral sectors transform differently, there are two sets of vielbeins and gamma matrices. We obtained the general form of the local Lorentz transformation in spinorial representation ${}_a\Omega$ connecting two chiral sectors. In order to work with properly defined variables and
background fields, fermions with bar index are multiplied by matrix ${}_a\Omega$. 

The T-dual transformation of dilaton field $\Phi(x^i,\theta^\alpha,\bar\theta^\alpha)$ is treated within quantum formalism. In this paper, using the matrices $\tilde\Pi_{+ab}$ and ${}_a \tilde\Pi_{+}^{ab}$, we obtained
the general expression for T-dual dilaton field (\ref{eq:dsh}).

Consequently, in this article we performed Buscher simultaneous T-dualization of type II superstring in pure spinor formulation and found the general form of the T-dual transformation laws and full expressions for T-dual background
fields.


\begin{thebibliography}{99}
\bibitem{mtheory} E. Witten, {\it Nucl. Phys.} {\bf B 443} (1995) 85; T. Banks, W. Fischler, S.H. Shenker, L. Susskind, {\it Phys. Rev.} {\bf D 55} (1997) 5112;  
M. J. Duff, {\it Int.J.Mod.Phys.} {\bf A11} (1996) 5623-5642; M. J. Duff, J. T. Liu, R. Minasian, {\it Nucl. Phys.} {\bf B452} (1995) 261-282; M. J. Duff, P. S. Howe, T. Inami, K. S. Stelle, {\it Phys.Lett.} {\bf B191} (1985) 70; 
P. Horava, E. Witten, {\it Nucl.Phys.} {\bf B 460} (1996) 506-524; 
C. M. Hull, P. K. Townsend, {\it Nucl.Phys.} {\bf B 438} (1995) 109-137; 
C. M. Hull, R. R. Khuri, {\it Nucl. Phys.} {\bf B 536} (1999) 219-244; R. Dijkgraaf, E. Verlinde, H. Verlinde, {\it Nucl. Phys.} {\bf B 500} (1997) 43-61.



\bibitem{IIAIIB} A. Kehagias, {\it Phys.Lett.} {\bf B 377} (1996) 241-244; M. Cvetic, H. Lu, C.N. Pope, K.S. Stelle, {\it Nucl.Phys.} {\bf B 573} (2000) 149-176; 
I. Jeon, K. Lee, J.-H. Park, Y. Suh, {\it Phys. Lett.} {\bf B 723} (2013) 245-250.

\bibitem{BNBSCBC} B. Nikolic, B. Sazdovic, {\it Eur.Phys.J.} {\bf C 77} (2017) 197.

\bibitem{GR} A. Giveon and M. Ro\v{c}ek, {\it Nucl. Phys.} {\bf B 421} (1994) 173.
\bibitem{H} S. F. Hassan, {\it Nucl. Phys.} {\bf B 568} (2000) 145.
\bibitem{BPT} R. Benichou, G. Policastro and J. Troost, {\it Phys. Lett.} {\bf B661} (2008) 129.
\bibitem{erik} E. Plauschinn, {\it Nucl.Phys.} {\bf B893} (2015) 257-286.
\bibitem{verteks} P. A. Grassi, L. Tamassia, {\it JHEP} {\bf 07} (2004) 071.


\bibitem{timelike} C. M. Hull, {\it JHEP} {\bf 07} (1998) 021.




\bibitem{B} T. Buscher, {\it Phys. Lett.} {\bf B 194} (1987) 59; {\bf 201} (1988) 466.
\bibitem{S} J. Polchinski, {\it String theory - Volume II}, Cambridge University Press,
1998; B. Zwiebach, {\it A First Course in String Theory}, Cambridge University Press, 2004; 
K. Becker, M. Becker and J. Schwarz {\it String Theory and M-Theory: A Modern Introduction} Cambridge University Press, 2007.

\bibitem{RV} M. Ro\v{c}ek and E. Verlinde, {\it Nucl.Phys.} {\bf B 373} (1992) 630.

\bibitem{GPR} A. Giveon, M. Porrati and E. Rabinovici, {\it Phys. Rep.} {\bf 244} (1994) 77.

\bibitem{AABL} E. Alvarez, L. Alvarez-Gaume, J. Barbon and Y. Lozano, {\it Nucl. Phys.} {\bf B 415} (1994) 71.

\bibitem{open} B. Sazdovi\'c, {\it Eur. Phys. J.} {\bf C77} (2017) 634; F. Cordonier-Tello, D. Lust, E. Plauschinn, Open-string T-duality and applications to non-geometric backgrounds, arXiv:1806.01308.

\bibitem{berko} N. Berkovits, hep-th/0209059; P. A. ~Grassi, G. Policastro and P. van Nieuwenhuizen, {\it JHEP} {\bf 10} (2002) 054; P. ~A. ~Grassi, G. ~Policastro and P. ~van ~Nieuwenhuizen, {\it JHEP} {\bf 11} (2002) 004; P. ~A. ~Grassi, G. ~Policastro and P. ~van ~Nieuwenhuizen, {\it Adv. Theor. Math. Phys.} {\bf 7} (2003) 499; P. ~A. ~Grassi, G. ~Policastro and P. ~van ~Nieuwenhuizen, {\it Phys. Lett.} {\bf B553} (2003) 96.

\bibitem{pspin} N. Berkovits, {\it JHEP} {\bf 07} (2015) 091; N. Berkovits, H. Gomez, arxiv: 1711.09966; M. Cederwall, {\it Springer Proc. Phys.} {\bf 153} (2014) 61-93.

\bibitem{susyNC} J. ~de ~Boer, P. ~A. ~Grassi and P. ~van ~Nieuwenhuizen, {\it Phys. Lett.} {\bf B574} (2003) 98.


\bibitem{NPBref} N. ~Berkovits and P. Howe, {\it Nucl. Phys.} {\bf B635} (2002) 75.


\end{thebibliography}
\end{document}